\documentclass[%
 aip,
cp,  
 amsmath,amssymb,
 reprint,%
]{revtex4-2}
\usepackage{color}
\usepackage{graphicx}
\usepackage{dcolumn}
\usepackage{bm}
\usepackage[utf8]{inputenc}
\usepackage[T1]{fontenc}
\usepackage{mathptmx} 
\begin{document}
\title{Mitigation Strategies for ${}^{42}$Ar/${}^{42}$K Background Reduction using Encapsulation with Ultra-Pure Plastic for the LEGEND Experiment}
\author{M. Ibrahim Mirza for the LEGEND Collaboration$^{1,}$} 
\email[Corresponding author: ]{mmirza3@vols.utk.edu}
\affiliation{
  $^1$Department of Physics and Astronomy, The University of Tennessee, Knoxville, TN 37996, United States
 }
\begin{abstract}Neutrinoless double-beta (0$\nu\beta\beta$) decay is the most compelling approach to determine the Majorana nature of neutrino and measure effective Majorana neutrino mass. The LEGEND collaboration is aiming to look for 0$\nu\beta\beta$ decay of ${}^{76}$Ge with unprecedented sensitivity. If underground-sourced argon is not available, the cosmogenically-induced isotope ${}^{42}$Ar and its decay progeny ${}^{42}$K in the liquid argon active veto could create a challenging background for the 0$\nu\beta\beta$ signal. We are studying methodologies to mitigate the ${}^{42}$K background. In order to achieve this, encapsulation of germanium detectors with 3D-printed technologies using low background material are currently under investigation. Simulation results of Poly(ethylene 2,6- naphthalate) (PEN) encapsulation of germanium detectors and plans to study other potential materials are presented.
\end{abstract}
\maketitle
\section{Introduction}
The Large Enriched Germanium Experiment for Neutrinoless $\beta\beta$ Decay (LEGEND) is a ultra low radioactive background experiment designed to detect 0$\nu\beta\beta$ decay signal in a ${}^{76}$Ge (Q$_{\beta\beta}$ = 2039 keV) \cite{legend}. Performance parameters of the 200 kg and 1000 kg phases of the LEGEND are listed in table \ref{tab:par}. Backgrounds in the region of interest (ROI) can be produced by detector components such as the detector mounts, front end electronics, cables, optical fibers, re-entrant tube, liquid argon, cryostat, among other sources.
\begin{table}[ht!]
\centering
\begin{tabular}{ |l|c| c |}
  \hline
\textbf{\footnotesize{Experiment}}  & \textbf{\footnotesize{LEGEND-200}} & \textbf{\footnotesize{LEGEND-1000}} \\
  \hline
\footnotesize{Status} &  \footnotesize{In commissioning} & \footnotesize{Planned}\\ 
   \hline
 \footnotesize{Location} &  \footnotesize{LNGS} & \footnotesize{LNGS / SNOLAB}\\ 
  \hline
 \footnotesize{Isotope} & \footnotesize{${}^{76}$Ge} & \footnotesize{${}^{76}$Ge}\\
  \hline
   \footnotesize{Total mass}& \footnotesize{200 kg} & \footnotesize{1000 kg} \\
      \hline
     \footnotesize{Exposure} & \footnotesize{1 t yr} & \footnotesize{10 t yr} \\
      \hline
      \footnotesize{Background Index} & \footnotesize{2$\times$10$^{-4}$ cts/(keV kg yr)} & \footnotesize{< 1$\times$10$^{-5}$ cts/(keV kg yr)}\\
      \hline
   \footnotesize{Discovery Sensitivity} & \footnotesize{10$^{27}$ yr} & \footnotesize{1.3 $\times$ 10$^{28}$ yr} \\
      \hline
    \footnotesize{Live time}& \footnotesize{5 yr} & \footnotesize{10 yr} \\
      \hline
      \footnotesize{Majorana neutrino mass} & \footnotesize{< 34 - 78 meV} & \footnotesize{< 9 - 21 meV}\\
      \footnotesize{(discovery sensitivity)}& & \\
      \hline
      \end{tabular}
\caption{The specifications for the two experimental phases of the LEGEND program.}
\label{tab:par}
\end{table}
\\ The work presented here is focused on the concepts and techniques to mitigate the background caused by argon, i.e. the small concentration of ${}^{42}$Ar and its daughter isotope ${}^{42}$K in atmospheric argon.
\subsection{${}^{42}$Ar / ${}^{42}$K BACKGROUND IN LEGEND}
\begin{figure}[ht!]
\centering
\begin{minipage}[t]{0.3\textwidth}
\includegraphics[width=4.5cm]{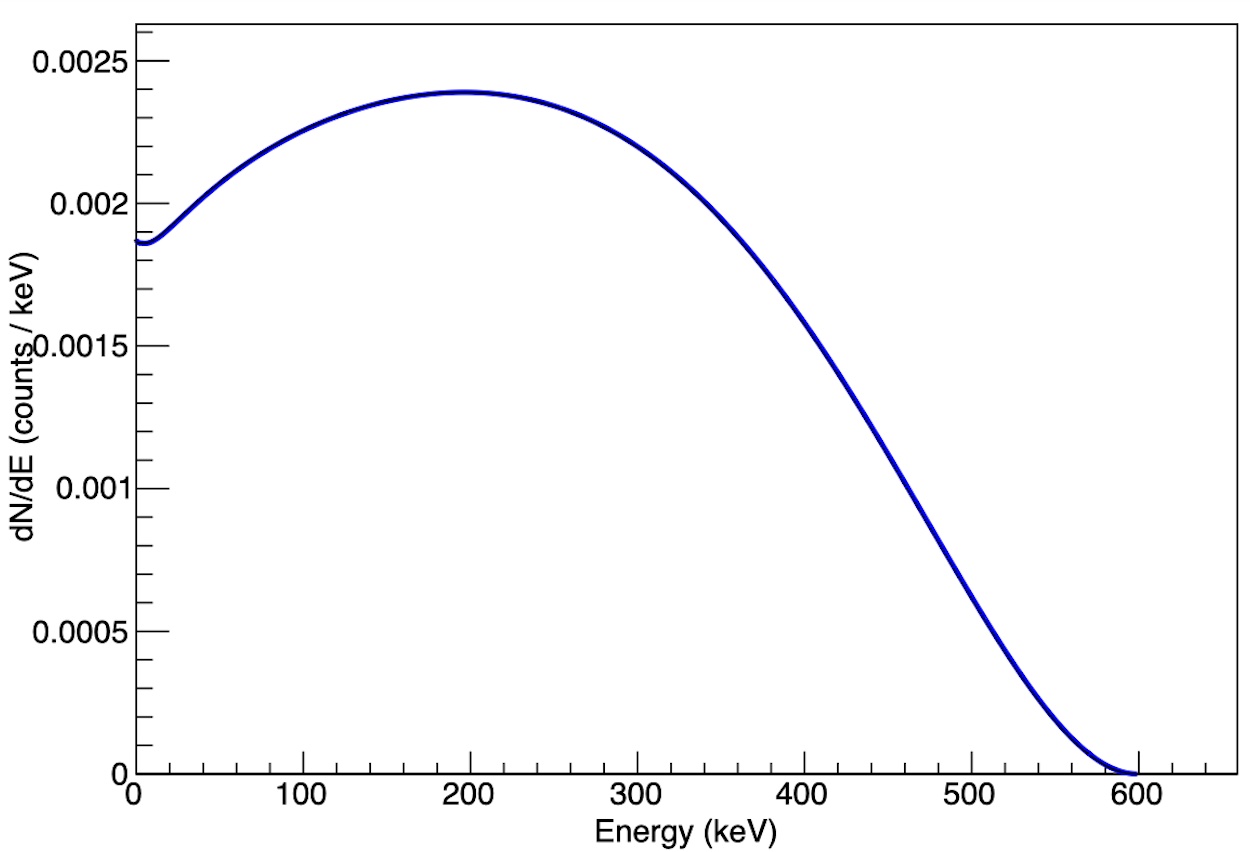}
\end{minipage}
\hspace{1cm}
\begin{minipage}[t]{0.3\textwidth}
\includegraphics[width=4.5cm]{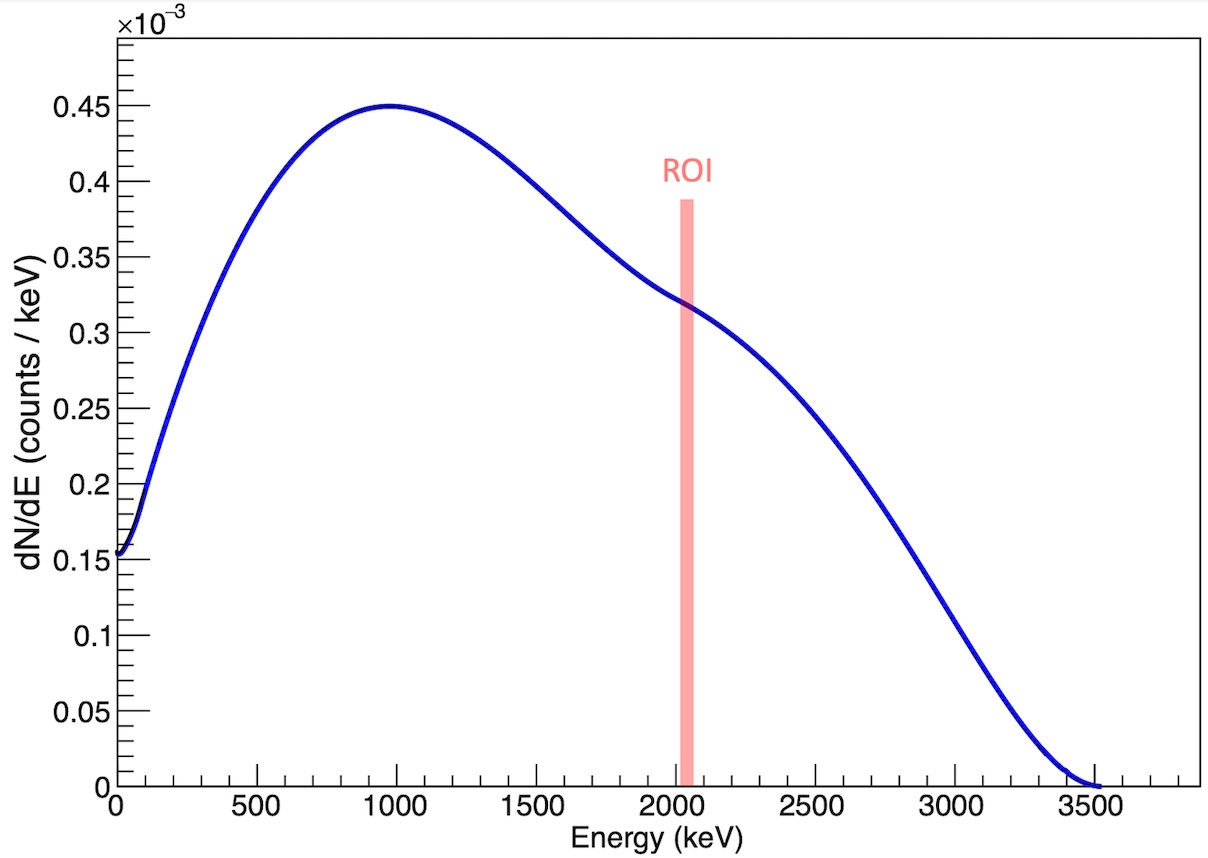}
\end{minipage}
\caption{Left: The energy distribution of ${}^{42}$Ar $\beta^-$ decay. Right: The energy distribution of ${}^{42}$K $\beta^-$ decay. The red band depicts the ROI. Data is taken from Ref. \cite{nds}.}
\label{fig:arg}
\end{figure}

Ge detectors are immersed in liquid argon in the LEGEND cryostat. For LEGEND, liquid argon serves as (i) coolant: because Ge detectors cannot be fully depleted at room temperature, therefore, they must operate at cryogenic temperature, (ii) veto for background events: liquid argon is a scintillator, therefore, by detecting scintillation light near Ge detectors, many background signals can be effectively suppressed. However, natural argon contains a small amount of ${}^{42}$Ar, which in turn produces ${}^{42}$K, a source of background. GERDA Phase II \cite{gerda} and DEAP-3600 \cite{deap} reported the presence of ${}^{42}$K in liquid argon. nEXO \cite{nexo} also reported the study of ${}^{42}$K in liquid xenon. 
\\ \textbf{Origin of ${}^{42}$Ar \& ${}^{42}$K in a atmospheric argon}: Gaseous argon in the upper atmosphere is constantly bombarded with cosmic rays. When a cosmogenically produced $\alpha$ particle interacts with atmospheric argon, it produces a small concentration of ${}^{42}$Ar via the reaction, ${}$ ${}^{40}$Ar + $\alpha$  $\rightarrow $ ${}^{42}$Ar + p + p. Liquid argon is normally extracted from the atmosphere, so ${}^{42}$Ar is present. ${}^{42}$Ar is a long lived radioactive isotope, with a half-life of 32.9 y. It undergoes $\beta^-$ decay via, ${}$ ${}^{42}$Ar $\rightarrow$ ${}^{42}$K$^+$ + $\beta^-$ + $\overline{\nu_e}$, with Q$_{\beta}$ = 0.599 MeV. The $\beta$ decay spectrum is shown in figure $\ref{fig:arg}$ (left). Since Q$_{\beta}$ $<$ Q$_{\beta\beta}$, ${}^{42}$Ar does not cause a background at the ROI for the LEGEND experiment. 
\\ \indent Additionally, the daughter isotope of ${}^{42}$Ar is ${}^{42}$K, which is also radioactive, with a short half-life of 12.36 h. It undergoes $\beta^-$ decay via, ${}$ ${}^{42}$K$^+$ $\rightarrow$ ${}^{42}$Ca$^{++}$ + $\beta^-$ + $\overline{\nu_e}$, with Q$_{\beta}$ = 3.525 MeV. The $\beta$ decay spectrum is shown in figure $\ref{fig:arg}$ (right). Since Q$_{\beta}$ $>$ Q$_{\beta\beta}$ the $\beta$ emitted in the ${}^{42}$K decay could cause a background at the ROI. 
\\ \indent ${}^{42}$K$^+$ can be attracted by the electric field of the detectors and as a result, accumulate on the Ge surface. Thermally driven convection in liquid argon also affects the dynamics of both neutral (${}^{42}$Ar) and charged (${}^{42}$K$^+$, $\beta^-$) particles. Backgrounds from ${}^{42}$K decay are expected to comprise 30-40$\%$ of LEGEND backgrounds in both phases of the experiment. Therefore, it is essential to reduce the concentration of ${}^{42}$Ar in the argon veto, or otherwise mitigate this background.
\\\textbf{Underground argon}: The most promising approach to dramatically reduce ${}^{42}$Ar backgrounds is the use of underground argon (UGLAr) i.e. argon extracted from deep underground sources. UGLAr is exposed to a much lower flux of cosmic particles, with no mechanism for ${}^{42}$Ar production by high-energy $\alpha$ particles. We plan to coordinate with the DarkSide-20k effort \cite{Aalseth:2017fik} to use the URANIA plant to produce UGLAr extracted from CO$_2$ wells in Colorado, which will then be transferred to the ARIA plant or other facilities for additional chemical purification. The LEGEND-1000 project is in negotiations to obtain approximately 18 tonnes of UGLAr. However, given the high risk associated with this background, the collaboration is investigating additional strategies to suppress the ${}^{42}$Ar background. 
\\ \textbf{Alternative methods to mitigate ${}^{42}$Ar background}: Several techniques to suppress the ${}^{42}$Ar background from argon are being studied by the collaboration: (i) encapsulation of Ge detectors, (ii) slightly depleting atmospheric sourced argon of ${}^{42}$Ar, (iii) enclosing the detector strings with ultra-pure nylon shroud with reduced argon volume as in LEGEND-200, (iv) fabricating Ge detectors with thicker n+ dead layers, (v) doping liquid argon with non-quenching impurities to neutralize ${}^{42}$K ions, (vi) better pulse-shape discrimination of $\beta$-induced events.
\\ Many of these techniques can be used to reduce backgrounds whether or not UGLAr is available. Among the above options in this article we focus on the encapsulation of the Ge detectors with low radioactivity scintillating material.
\section{Encapsulation of the Germanium Detector}
GERDA phase II data shows the viability of encapsulation for ${}^{42}$K background suppression \cite{gerda,gerda2}. GERDA used the transparent nylon mini shrouds to limit the volume of argon that could drift into the vicinity of the Ge detectors. However, to completely eliminate the possibility for ions to drift toward the Ge detectors, we need to provide a complete encapsulation. 392 Ge detectors with a total mass of 1000 kg of ${}^{76}$Ge will be deployed in the LEGEND-1000 cryostat. All of the Ge detectors will have different sizes and shapes, so capsules would need to be custom built for every detector. 3D printing is a viable solution for encapsulation to meet these demands and Oak Ridge National Laboratory (ORNL) facilities are equipped with state of the art 3D printers well suited for R\&D studies. Custom-produced PEN is another material considered for encapsulation.
\subsection{Simulation Results of Encapsulation}
\begin{figure}[ht!]
\centering
\begin{minipage}[t]{0.25\textwidth}
\includegraphics[width=3.1cm]{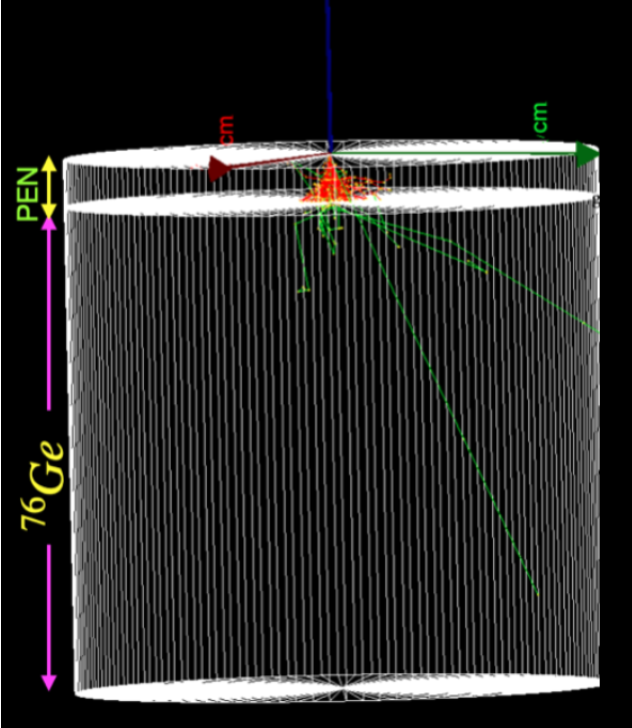}
\end{minipage}
\hspace{0.1cm}
\begin{minipage}[t]{0.25\textwidth}
\includegraphics[width=3.8cm]{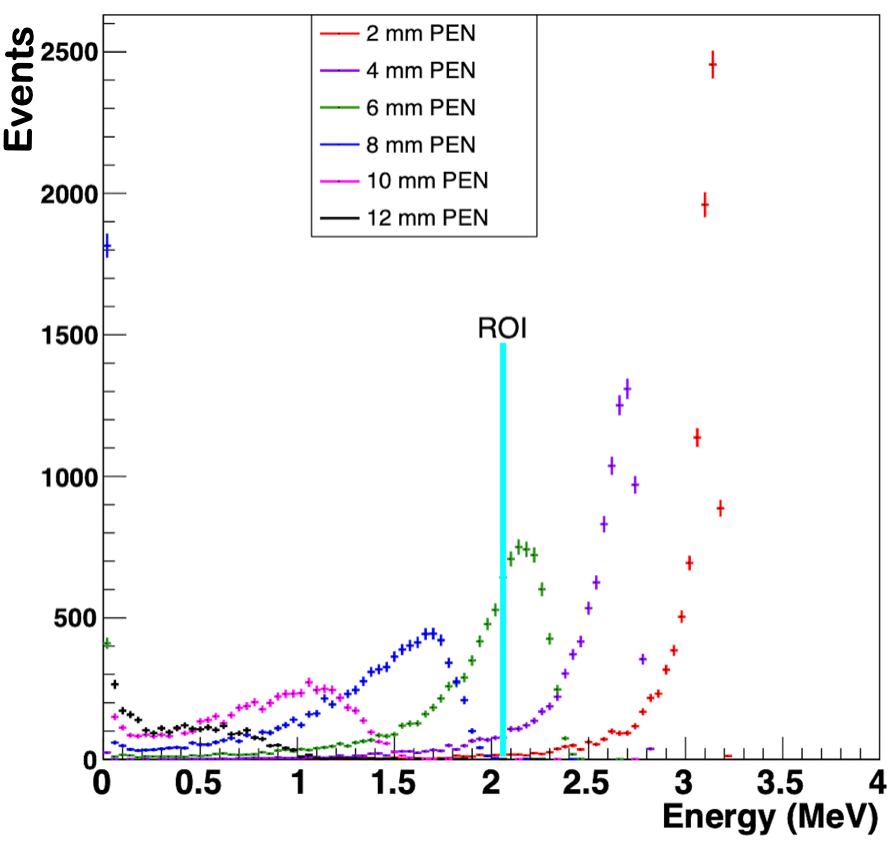}
\end{minipage}
\hspace{0.1cm}
\begin{minipage}[t]{0.25\textwidth}
\includegraphics[width=3.8cm]{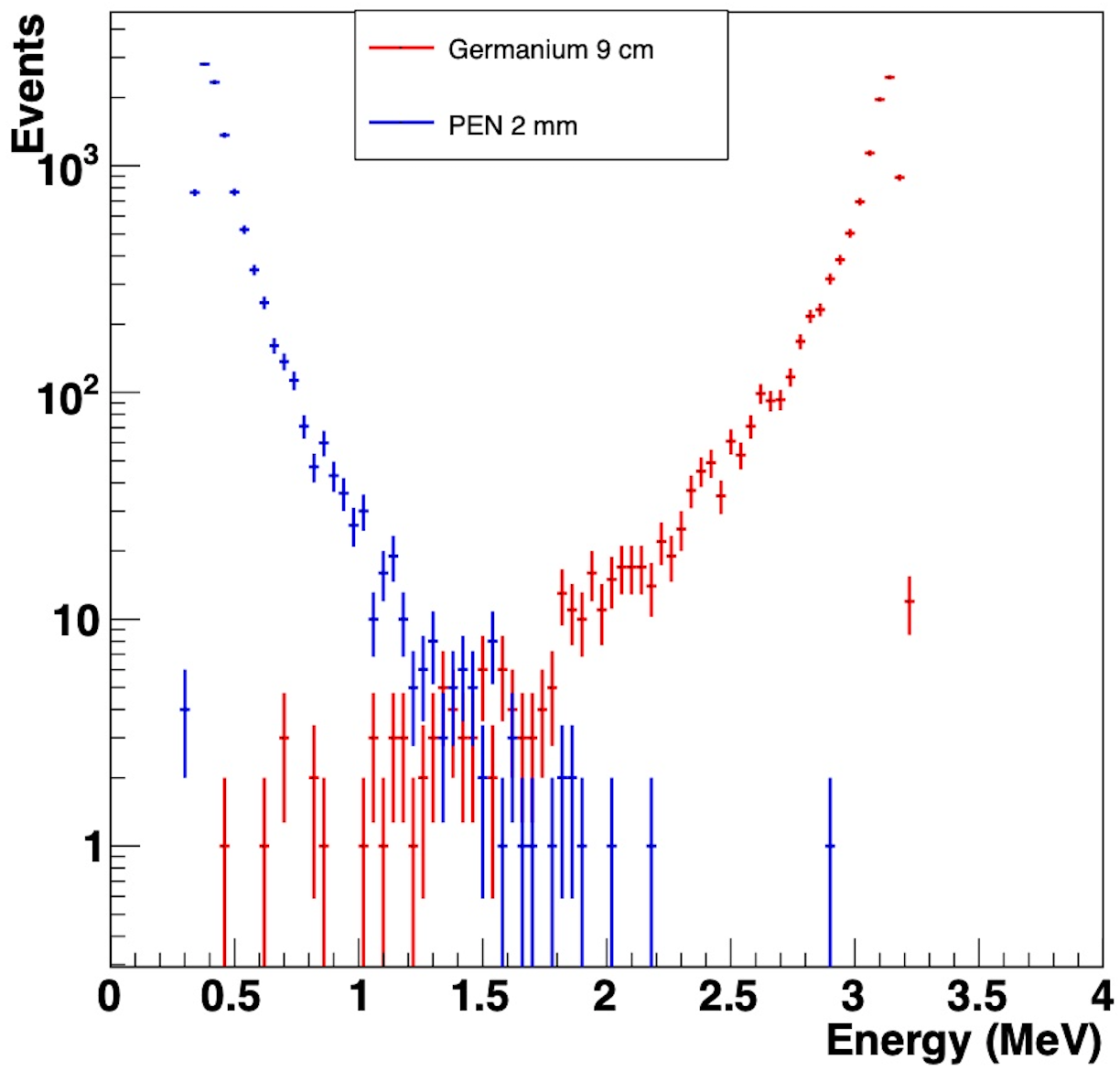}
\end{minipage}
\caption{Left: simulation of the detector geometry, with red representing the trajectories of the electrons in PEN and green representing the resulting gammas trajectories in Ge. Middle: simulation results of energy deposition in Ge, where the blue band depicts the ROI. Right: simulated energy deposition in Ge (red) and PEN (blue).}
\label{fig:fed}
\end{figure}

Poly(ethylene 2,6- naphthalate) (PEN) is a transparent plastic scintillator that has proven its capability as a wavelength shifter from vacuum ultra violet (VUV) into the blue region of the visible spectrum \cite{pen,pen2}. PEN have a same mechanical stability as copper at cryogenic temperature and LEGEND-200 employs PEN as a baseplate to support each Ge detector. 
\\ \indent Due to it's favorable characteristics, we consider a encapsulation design made entirely of PEN. Using GEANT4 \cite{gean}, we simulate the geometry of PEN encapsulated Ge. A layer of PEN is deposited over the top cross-section of the Ge detector and surrounded with liquid argon, as shown in figure \ref{fig:fed} (left). At the center of the PEN disk, a source of electrons with energy of 3.525 MeV, equal to Q$_{\beta}$ of ${}^{42}$K is simulated. We select the direction of the particle momenta to be towards the PEN. We calculated the energy deposition by the electrons in Ge when varying the PEN thickness, as shown in figure \ref{fig:fed} (middle). The simulation shows that 8 mm of PEN is required to reduce the energy of 3.525 MeV electrons to below the ROI. This is a lot of passive material next to Ge detector and can create another source of radioactive background. This simulation, however, does not take into account the dead layer present on the surface of the Ge detectors (1-2 mm thick) or the role of pulse-shape rejection of near-surface events. Figure \ref{fig:fed} (right) shows the energy deposited in a 9 cm thick Ge detector and 2 mm thick PEN plate, from the incident electron. From this simulation we can conclude that 2 mm of scintillation material next to the Ge detector can efficiently tag products of ${}^{42}$K decay. We show this simulation as an illustrative example and not necessarily a viable encapsulation design.
\section{R\&D for Encapsulation}
R\&D for encapsulation of Ge detectors is currently ongoing at ORNL. The selection of material for encapsulation is under investigation. An ideal material for encapsulation should be: (i) radiopure, (ii) 3D printable, (iii) transparent, (iv) scintillating, (v) mechanically stable at cryogenic temperature, (vi) does not contaminate liquid argon by outgassing. In an attempt to choose the appropriate material for encapsulation, we have compiled a database of resins based on their physical properties. We are also considering tests of resins used for medical and dental purposes, as such resins have relatively high tensile strength compared to the resins used for engineering designs.
\subsection{Ongoing and upcoming activities for encapsulation}
\begin{figure}[ht!]
  \begin{minipage}[c]{0.5\textwidth}
    \includegraphics[width=3.3cm]{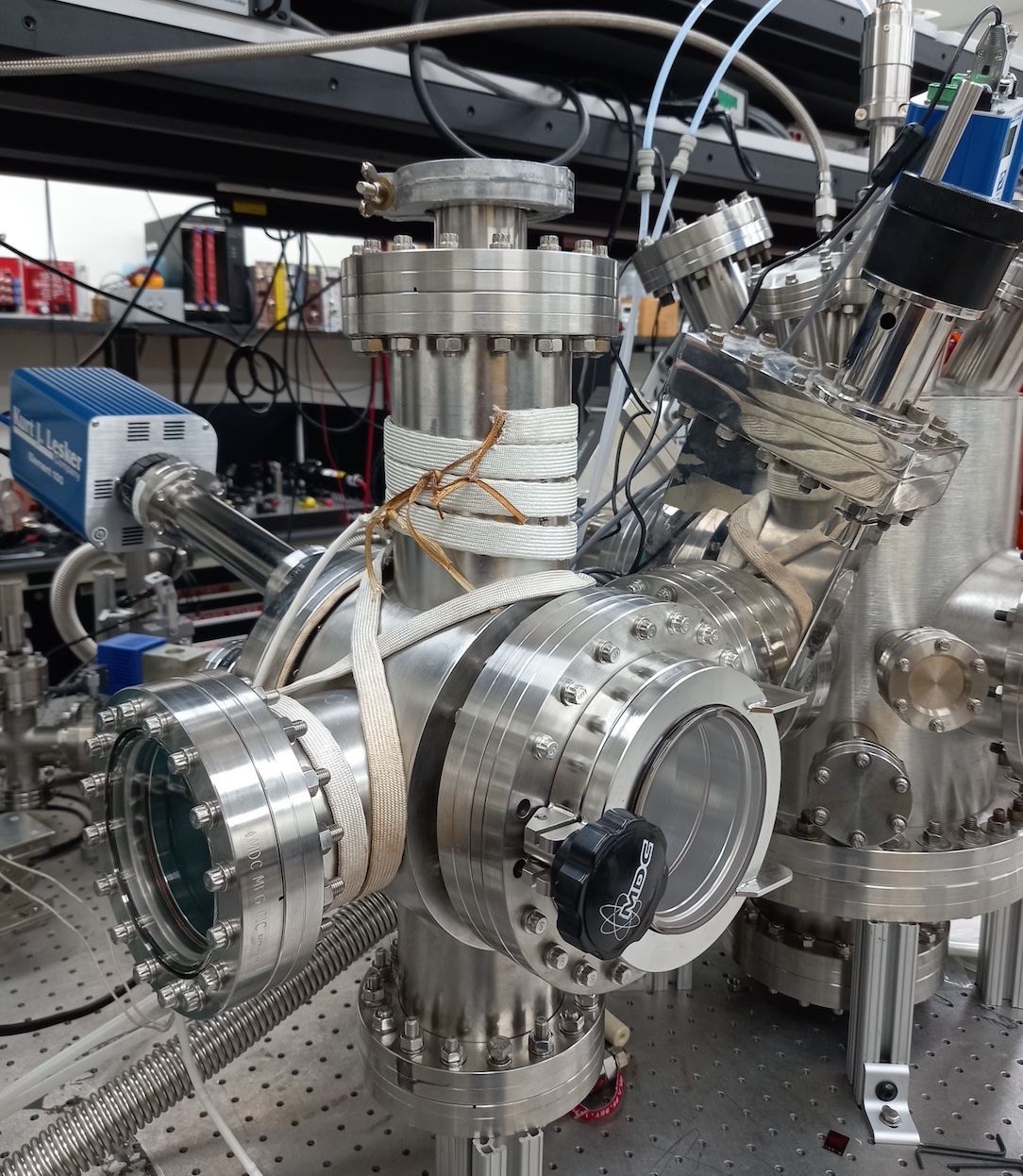}
 \end{minipage}\hfill
    \begin{minipage}[c]{0.5\textwidth}
    \caption{Recently assembled residual gas analyzer mounted on the outgassing vacuum chamber in the Physics Division at ORNL}
      \label{fig:drw}
  \end{minipage}
\end{figure}

 As part of our R\&D plan, we will perform the outgassing, radiopurity, optical, mechanical and cryogenic tests for the encapsulation material, which are described as follows:
\\ \textbf{Outgassing measurements:} The outgassing measurement identifies the molecular species present inside the encapsulation sample. UTK-ORNL jointly commissioned a setup for outgassing measurements at ORNL. We have assembled a residual gas analyzer (RGA) connected to a vacuum chamber, as shown in figure $\ref{fig:drw}$. The encapsulation materials are placed in the vacuum chamber. Prior to the outgassing measurements, a chamber pressure of around 5 $\times$ 10$^{-7}$ Torr is maintained by the combination of roughing and turbo pumps. Gaseous molecules present in the encapsulation sample will then move out and enter into the RGA. 
\\ \textbf{Radiopurity measurements:} In an attempt to meet LEGEND background goal, any supplementary materials used in the detector must have an ultra-low concentration of the U-238 and Th-232 decay chains and must not introduce impurities into the liquid argon, which can quench scintillation light. Hence the next step after the outgassing measurements would be an evaluation of radioactive impurities in the sample. We will perform ICP-MS and gamma ray spectrometry. The material having highest level of radiopurity will be selected as a potential candidate for encapsulation tests of Ge detectors.
\\\textbf{Optical measurements:} The encapsulation would leverage the liquid argon veto system for the LEGEND-1000 experiment. Therefore, it is essential to determine the optical properties of the encapsulation material. Every material under consideration will undergo optical testing. We intend to perform several measurements of the optical response of the encapsulation material, including photon yield, refractive index, attenuation length, peak emission wavelength and reflectivity. Existing equipment for optical reflectivity measurements at ORNL will be used, and we will establish an extensive optical test setup. 
\\\textbf{Mechanical measurements:} The encapsulation material will remain in the cryogenic environment of LEGEND-1000 for at least 10 years. Therefore, the structural integrity of the material is crucial and every material under considerations will also undergo extensive mechanical testing. We will test mechanical performance after the optical testing. Measurement of the tensile strength of each sample will be determined. For this purpose we intend to establish a mechanical testing setup.
\\\textbf{Cryogenic measurements:} Ge detectors will immersed in a liquid argon in the LEGEND-1000 cryostat. Therefore, the encapsulation samples will undergo cryogenic testing in liquid argon. We are currently planning to custom-build a cryostat with a VUV optical readout at ORNL. We will test the encapsulated Ge detector in an atmospheric argon, enriched with ${}^{42}$Ar. We are considering artificially producing ${}^{42}$Ar via, ${}^{40}$Ar(${}^{7}$Li$^{3+}$, $\alpha$p)${}^{42}$Ar reaction \cite{pel}.
\begin{acknowledgments}
We would like to acknowledge support by LEGEND Collaboration, including the ORNL LEGEND group and Physics Division. This work is supported by the U.S. DOE, and the NSF, the LANL, ORNL and LBNL LDRD programs; the European ERC and Horizon programs; the German DFG, BMBF, and MPG; the Italian INFN; 
the Polish NCN and MNiSW; the Czech MEYS; the Slovak SRDA; the Swiss SNF; the UK STFC; the Russian RFBR; the Canadian NSERC and CFI; the LNGS and SURF facilities.
\end{acknowledgments}
\bibliography{aipsamp}
\end{document}